\documentclass[]{spie}  

\usepackage{amsmath,amsfonts,amssymb}
\usepackage{graphicx}
\usepackage{IEEEtrantools}
\usepackage{indentfirst} 
\usepackage{graphicx}
\usepackage{subfigure}
\usepackage{makecell}
\usepackage{color}
\usepackage{cite}
\usepackage[colorlinks=true, allcolors=blue]{hyperref}

\title{Panoramic SETI: Overall focal plane electronics and timing and network protocols}

\author[a m]{Wei Liu}
\author[a b]{Dan Werthimer}
\author[a]{Ryan Lee}
\author[c]{Franklin Antonio}
\author[d]{Michael Aronson}
\author[e i]{Aaron Brown}
\author[f]{Frank Drake}
\author[g]{Andrew Howard}
\author[h]{Paul Horowitz}
\author[e i]{Jerome Maire}
\author[j]{Rick Raffanti}
\author[k]{Remington Stone}
\author[l]{Richard Treffers}
\author[e i]{Shelley A. Wright}

\affil[a]{Department of Astronomy, University of California Berkeley, USA}
\affil[b]{Space Sciences Laboratory, University of California Berkeley, CA, USA}
\affil[c]{Qualcomm Research Center, 5775 Morehouse Dr, San Diego, CA, USA}
\affil[d]{Electronic Packaging Man, Encinitas, CA, USA}
\affil[e]{Department of Physics, University of California San Diego, USA}
\affil[f]{SETI Institute, Mountain View, USA}
\affil[g]{Astronomy Department, California Institute of Technology, USA}
\affil[h]{Department of Physics, Harvard University, USA}
\affil[i]{Center for Astrophysics \& Space Sciences, University of California San Diego, USA}
\affil[j]{Techne Instruments, Berkeley, CA, USA}
\affil[k]{University of California Observatories, Lick Observatory, USA}
\affil[l]{Starman Systems, Alamo, USA}
\affil[m]{Institute of RF- \& OE-ICs, Southeast University, China}
\authorinfo{Further author information: (Send correspondence to W.L.)\\Wei Liu: E-mail: liuwei\_berkeley@berkeley.edu}

\begin{document} 
\maketitle 

\begin{abstract}
\indent The PANOSETI experiment is an all-sky, all-the-time visible search for nanosecond to millisecond time-scale transients. The experiment will deploy observatory domes at several sites, each dome containing $\sim$45 telescopes and covering $\sim$4,440 square degrees. Here we describe the focal-plane electronics for the visible wavelength telescopes, each of which contains a Mother Board and four Quadrant Boards. On each quadrant board, 256 silicon photomultiplier (SiPM) photon detectors are arranged to measure pulse heights to search for nanosecond time-scale pulses. To simultaneously examine pulse widths over a large range of time scales (nanoseconds to milliseconds), the instrument implements both a Continuous Imaging Mode (CI-Mode) and a Pulse Height Mode (PH-Mode). Precise timing is implemented in the gateware with the White Rabbit protocol.

\end{abstract}

\keywords{SETI, High speed Imaging, Wide Field Imaging, Pulse Height, Detector, SiPM, Instrumentation, White Rabbit, HDF5}

\section{INTRODUCTION}
	\label{sec:intro} 
\indent Modern SETI (the Search for Extra Terrestrial Intelligence) began in 1960, with Drake's ``Ozma'' search for radio signals from extraterrestrial civilizations, and a subsequent cascade of increasingly sophisticated and sensitive searches\cite{korpela2001seti}. While overlooked initially, it became evident that optical wavelengths could also be used for communication over modest interstellar distances\cite{townes1983wavelengths,schwartz1961interstellar}. Subsequently several groups engaged in optical  SETI\cite{howard2004search,horowitz2001targeted,werthimer2001berkeley,wright2001improved, wright2014near}, however these were unable to conduct observations with simultaneous large sky coverage\cite{2018Panoramic}. \\
\indent The PANOSETI experiment aims to search for fast transient signals (nano-second to seconds) by monitoring the entire observable sky during all observable time, which greatly expands the search-space of optical SETI. To cover a wide field of view, and to reduce the false alarm rate, the PANOSETI experiment will deploy telescopes at several sites. Each site will have roughly $\sim$45 telescopes in a dome, with each telescope pointing to a different direction, which is like a ``Fly's Eye Telescope Array.''\cite{siemion2011allen} Each telescope covers roughly 100 square degrees (10 by 10 degree field), so a dome equipped with $\sim$45 telescopes will cover $\sim$4440 square degrees simultaneously. If two or more sites detect a similar event, at roughly the same direction in the sky, and an arrival time difference consistent with a source at interstellar distance, then confidence is high that the event is interesting, and not from a false alarm, such as atmospheric Cherenkov radiation, or optical transients from satellite glints. \\
\indent Cross correlation is useful for detecting the interesting event with weak pulse transient signal at two or more sites. The cross correlation requires precision time stamping of the events at each site, which enables a crude measure of distance and discrimination of nearby phenomena. To identify and reject atmospheric transients, the differential timing accuracy of events should be nanosecond scale.   

\section{PANOSETI Hardware}
	\label{sec:title}
\begin{figure}[htb]
		\begin{center}
			\includegraphics[height=6cm]{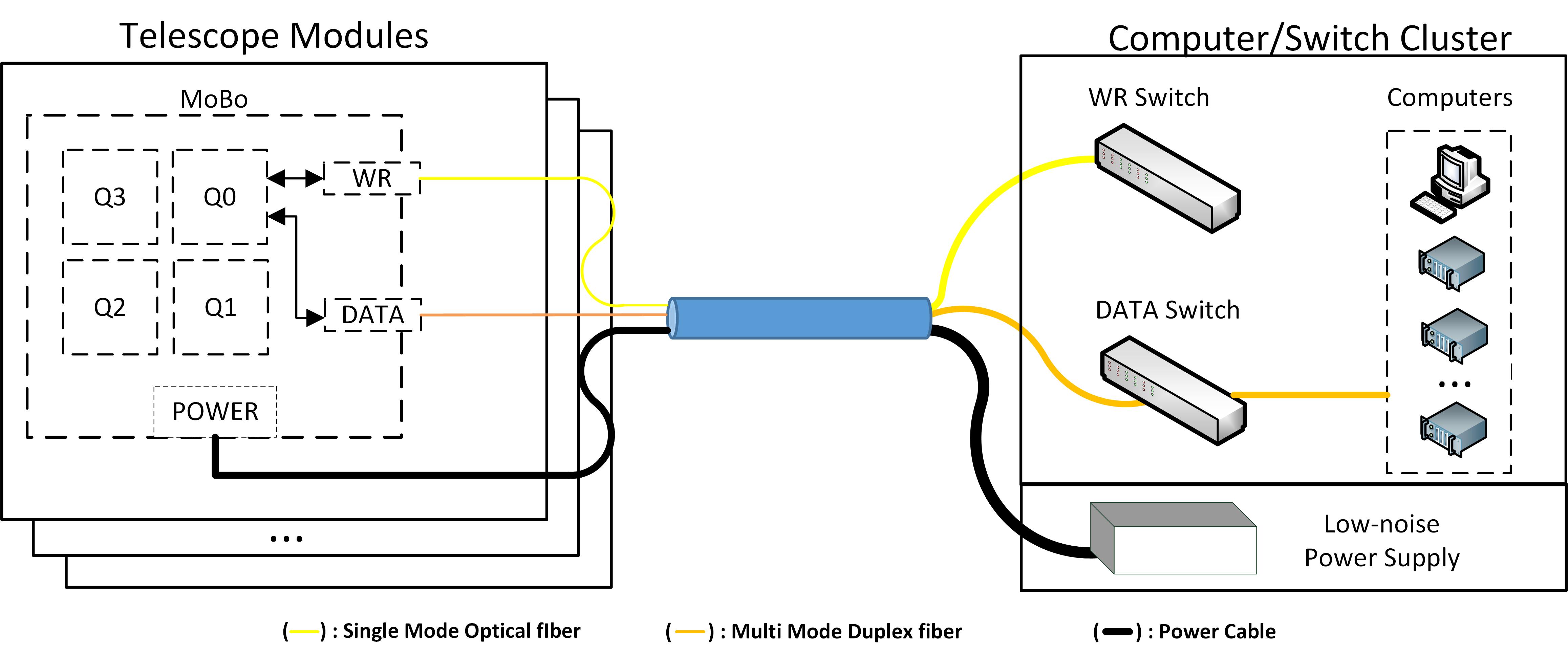}
		\end{center}
		\caption{
			\label{sch_of_system}
			Block Diagram of the Panoseti Electronic System.  WR is the white rabbit time and frequency distribution network.  
		}
	\end{figure}

\indent The PANOSETI search consists of a pair of domes, spaced $\sim$1\,km apart, each containing 45 telescope modules.  Each module has a 0.5\,m diameter Fresnel lens and a pixelated detector plane. The detector plane consists of a mother board (MoBo), on which four quadrant boards (Q0-Q3) are installed. The mother board also provides two SFP interfaces for scientific data and timing data transmission to switches. Quadrant boards are the key part of the whole electronic system, on which we have 256 high speed photon detectors and four 64-channel Maroc3 ASICs for amplifying, pulse shaping, and digitizing each detector's signal. Outboard rack-mounted power systems distribute 24\,Vdc and $-$70\,Vdc voltages to the telescope modules, where additional power conversion creates the lower supply voltages for the analog and digital circuits. The schematic of the electronic system is shown in Figure \ref{sch_of_system}. 

\subsection{Photon Detectors and Digitized ASIC}
\begin{figure}[htbp]
	\centering
	\includegraphics[height=3cm]{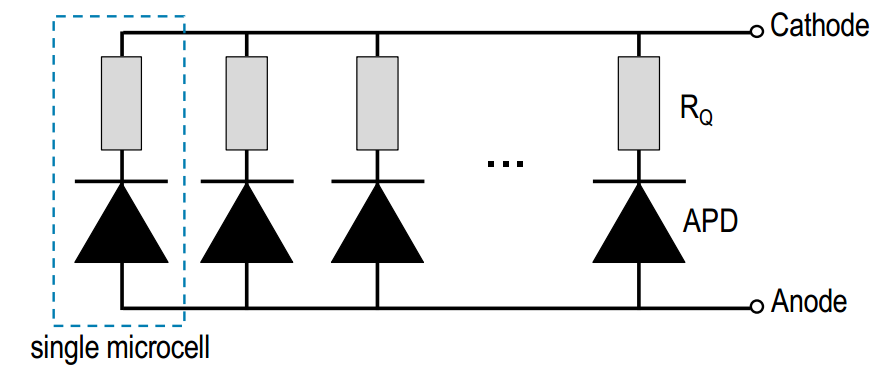}
	\caption{
		\label{microcell}
		Structure of a SiPM single pixel detector 
	}
	\end{figure}

\begin{figure}[htbp]
		\begin{center}
			\includegraphics[height=4cm]{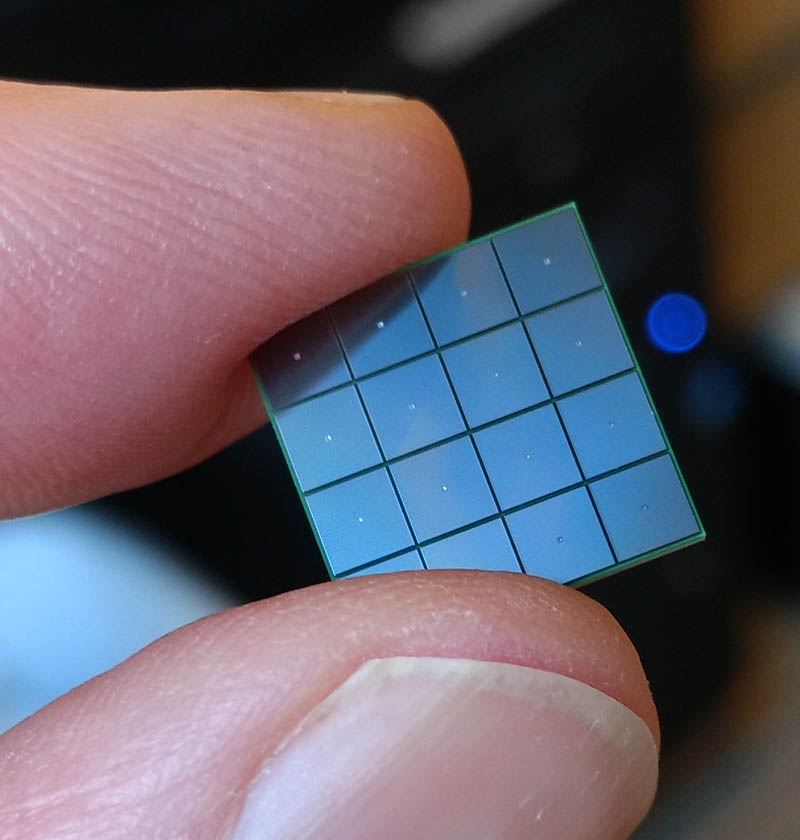}
		\end{center}
		\caption{
			\label{photon_detector}
			Photo of a 4x4 pixel SiPM array
		}
	\end{figure}

\indent The expensive photomultipliers of yesteryear have an inexpensive solid-state descendant, the ``silicon photomultiplier'' (SiPM), making possible a system that can cover a large portion of the sky simultaneously. Each pixel of an SiPM is itself an array of $\sim\!50\,\mu$m square microcells, each of which consists of a quenching resistor and avalanche photodiode. All $\sim$3,600 microcells in a single 3mm-by-3mm pixel are connected in parallel (Fig.~\ref{microcell}), producing a photon-sensitive detector with high efficiency ($\sim$25\%), high speed ($<$10\,ns), and low cost ($\sim$\$4 per pixel).  SiPMs are available in square arrays, e.g., 8$\times$8; we use four such arrays on each quadrant board, thus 256 pixels. Figure~\ref{photon_detector} shows a smaller 4$\times$4 SiPM array.\\

\indent To convert the analog signal to digital, we employ four ASICs, each with 64 analog channels. The millivolt-scale analog pulses from the 256 SiPM detectors on each quadrant board are conditioned by a set of four Weeroc Maroc3A ASICs: each ASIC has 64 channels of amplifiers, discriminators, and sample/holds, all packaged in a diminutive 12\,mm square package.\\

\subsection{Quadrant Board}
	\label{sec:quabo}
\begin{figure}[htbp]
		\begin{center}
			\includegraphics[height=8cm]{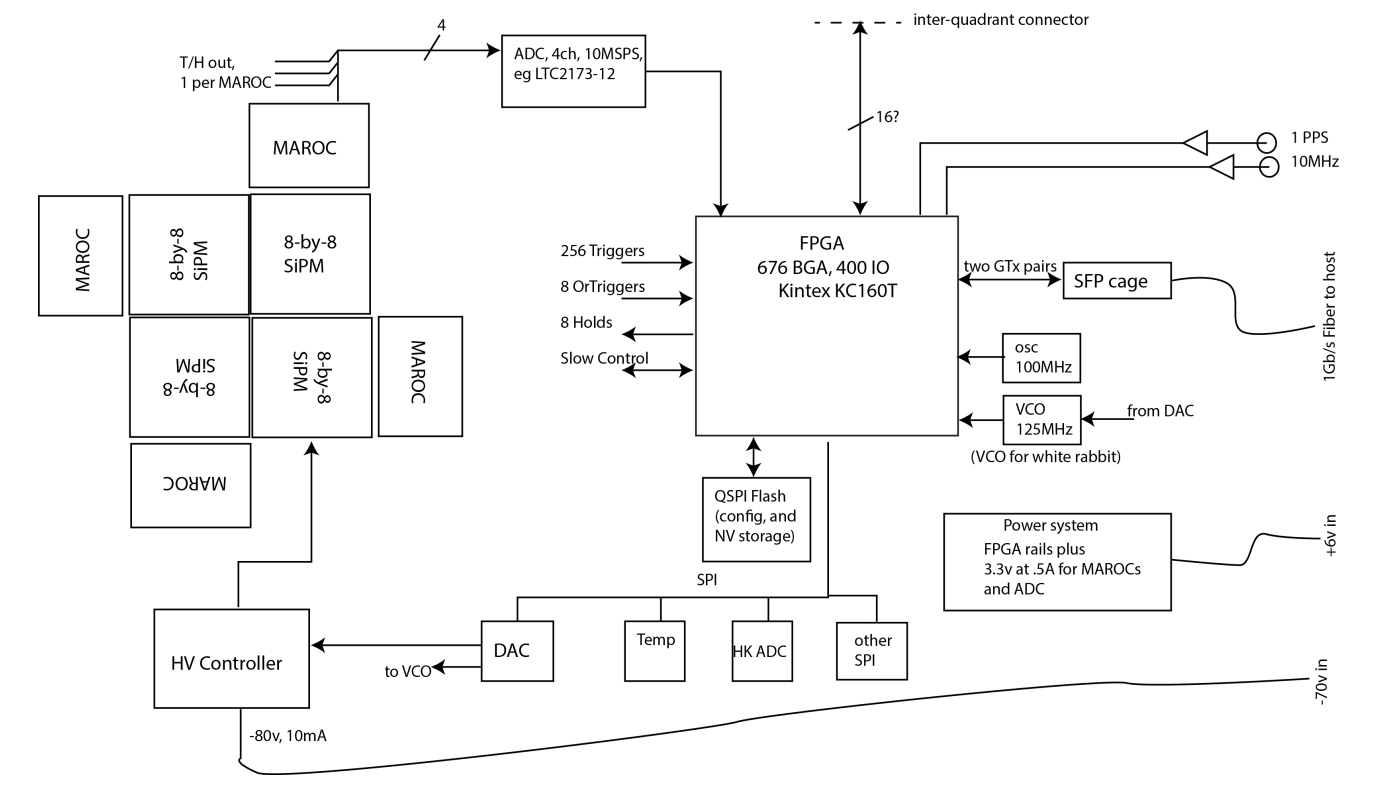}
		\end{center}
		\caption{
			\label{sch_of_quabo}
			Block Diagram of a 256 pixel Quadrant Board{\cite{2018Panoramic}} 
		}
	\end{figure}

\begin{figure}[htbp]
		\begin{center}
			\subfigure[Component Side]{
				\includegraphics[height=5cm]{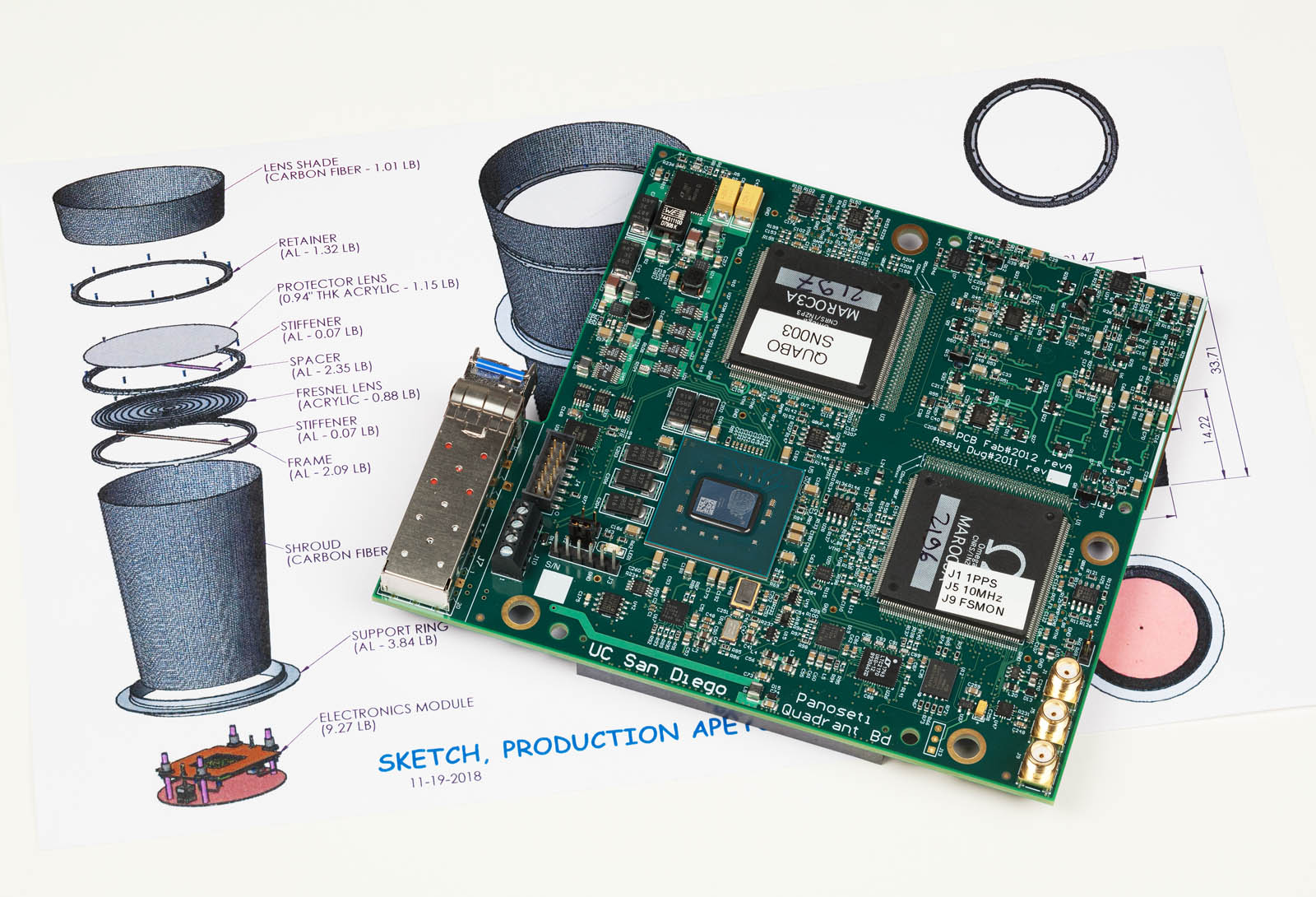}
			}
			\subfigure[Detector Side]{
				\includegraphics[height=5cm]{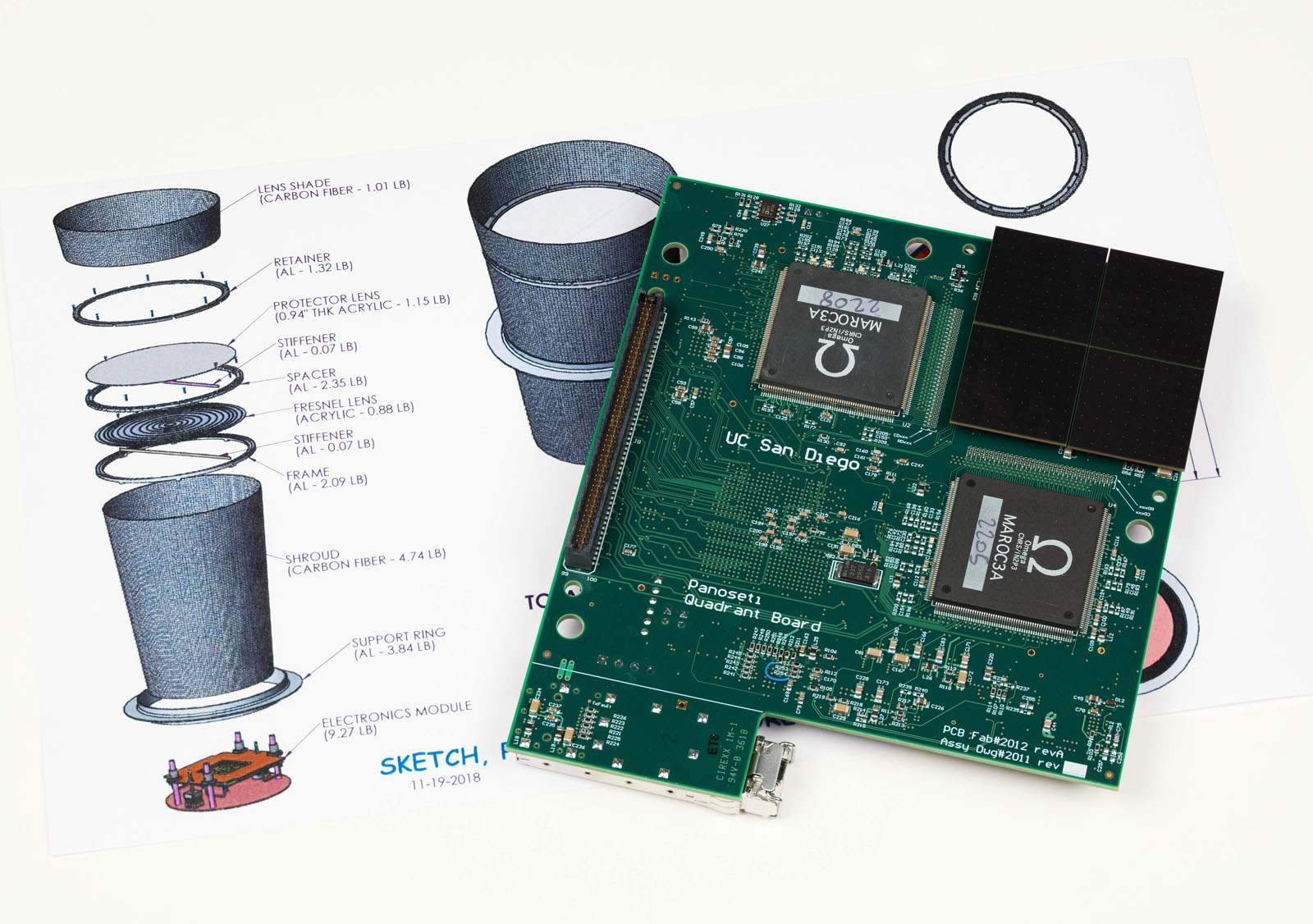}
			}
		\end{center}
		\caption{
			\label{photo_of_quabo}
			Photo of Quadrant Board.  Four Quadrant boards are deployed at each focal plane. 
		}
	\end{figure}

\indent The quadrant board is the most important part of the electronics. Each ``quabo'' has four 8-by-8 SiPM arrays and four Maroc3A chips (plus ADCs) for converting 256-pixel analog signals to digital quantities. It's based on a Xilinx XC7K160 Field Programmable Gate Array (FPGA) chip, which contains digital signal processing cores, a bank of 256 photon counters, application-specific logic circuits, and some other logic resources. There are also ADCs on the board, which are used for digitizing the amplified signal pulses, and for monitoring the status of the whole system, such as the temperature, voltages for SiPM, FPGA, and Maroc3A chips. Two 1Gb Ethernet ports implemented by GTX transceivers are connected to two SFP interfaces, so all the data are transmitted through fibers. 1\,GbEs is used to transmit the imaging, pulse-height data and housekeeping data to a bank of servers for real time processing, another is used for transmitting White Rabbit data, which is used for timing distribution and synchronization over Ethernet. A 125\,MHz VCO controlled by a DAC on the board is also used for White Rabbit. To synchronize the $45\times4\times2$=360 quadrant boards in the two domes, an external time/frequency reference is employed to provide 62.5\,MHz and 1\,PPS; the White Rabbit timing distribution technology is described below (\S\ref{sec:white_rabbit}). A QSPI flash memory on the board is used for storing bit files for booting up the FPGA, and some configuration parameters. The block diagram of a quadrant board is shown in Figure~\ref{sch_of_quabo}, and the photo of a quadrant board is shown in Figure~\ref{photo_of_quabo}.  \\

\subsection{Mother Board}
	\label{sec:mobo}
\begin{figure}[htbp]
	\centering
	\includegraphics[height=5cm]{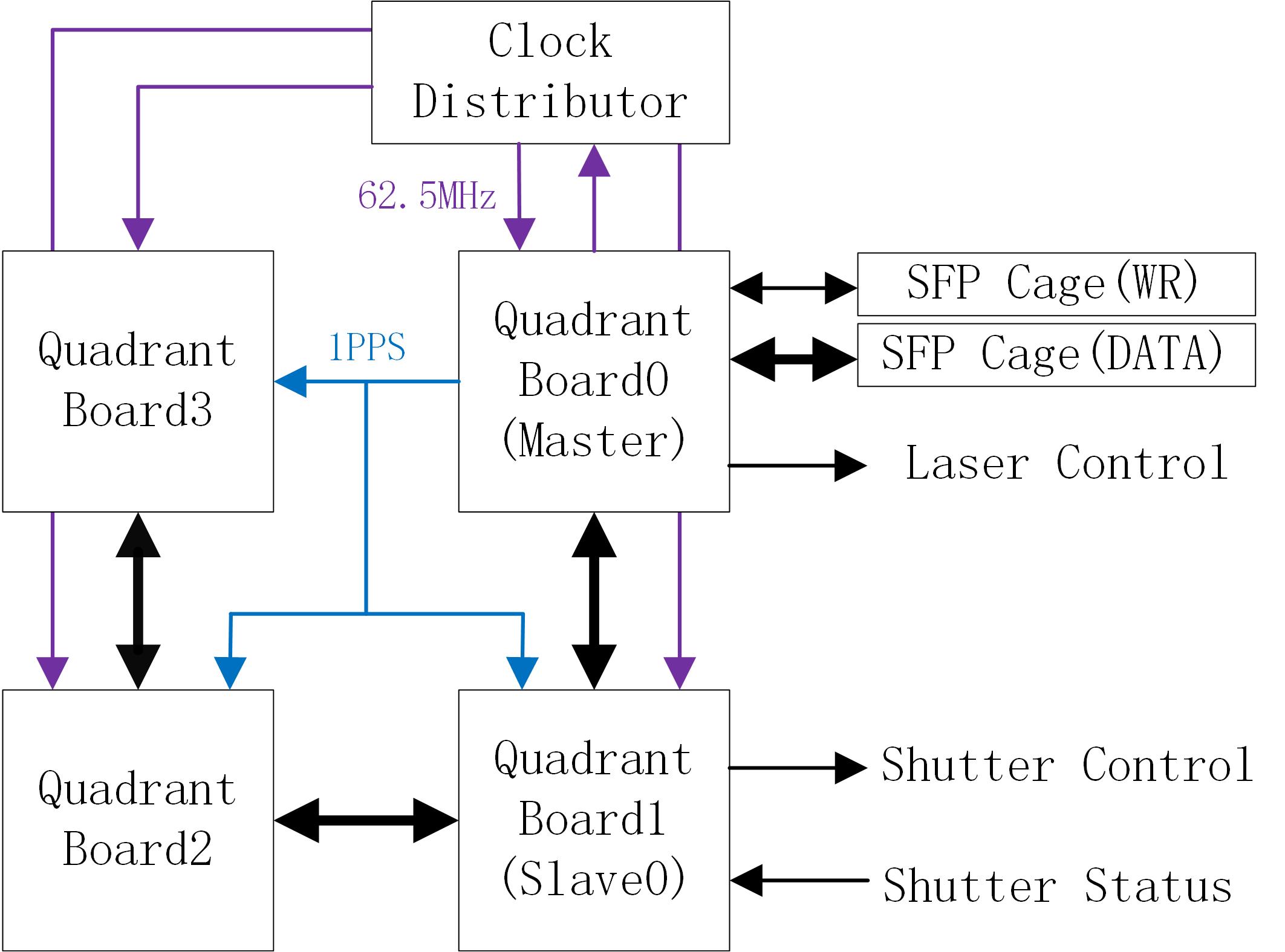}
	\caption{
		\label{sch_of_mobo}
		Block Diagram of PANOSETI's focal plane mother board
	}
\end{figure}

\begin{figure}[htbp]
	\begin{center}
		\subfigure[Under Side]{
			\includegraphics[height=5cm]{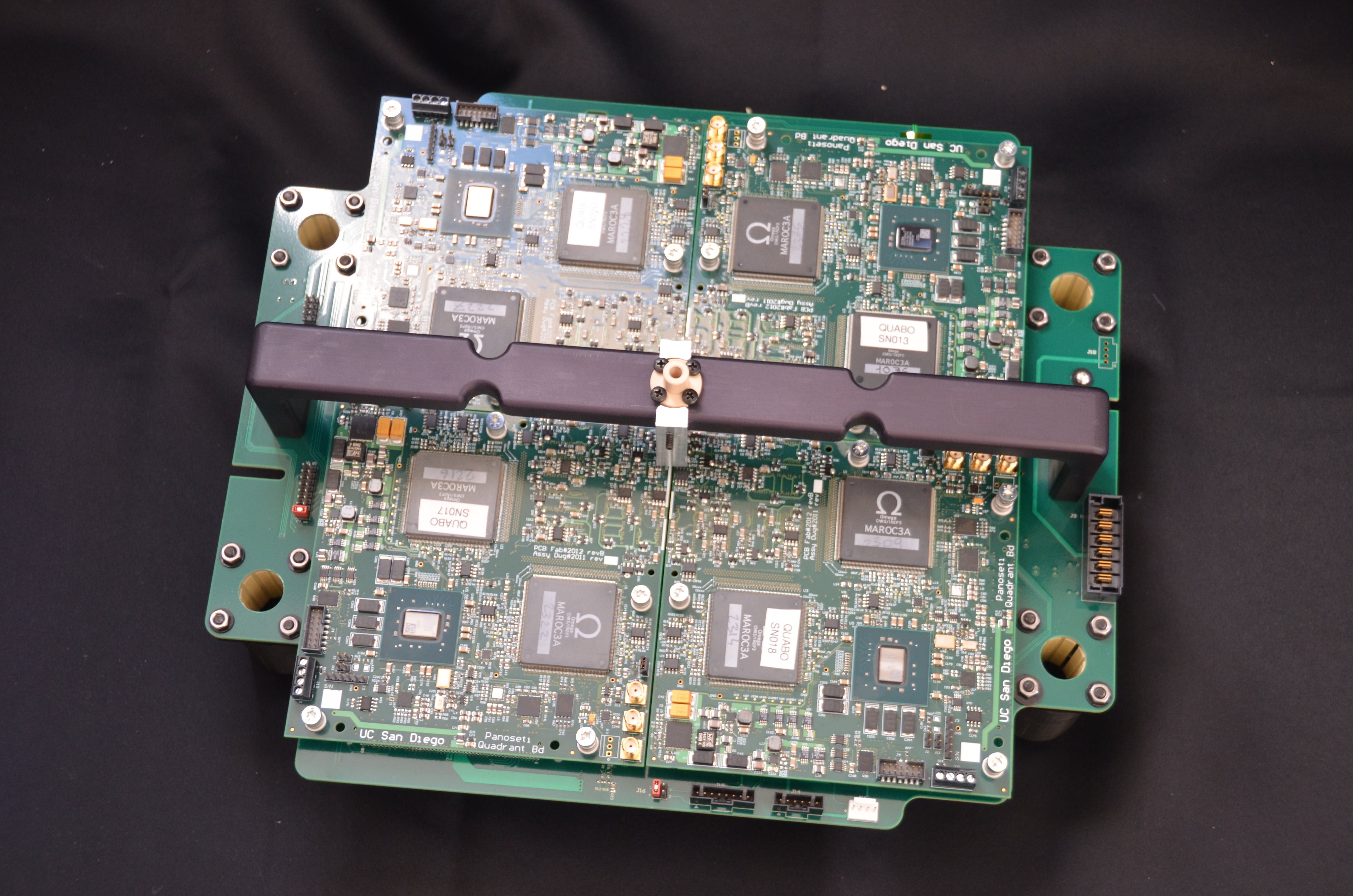}
		}
		\subfigure[Top Side]{
			\includegraphics[height=5cm]{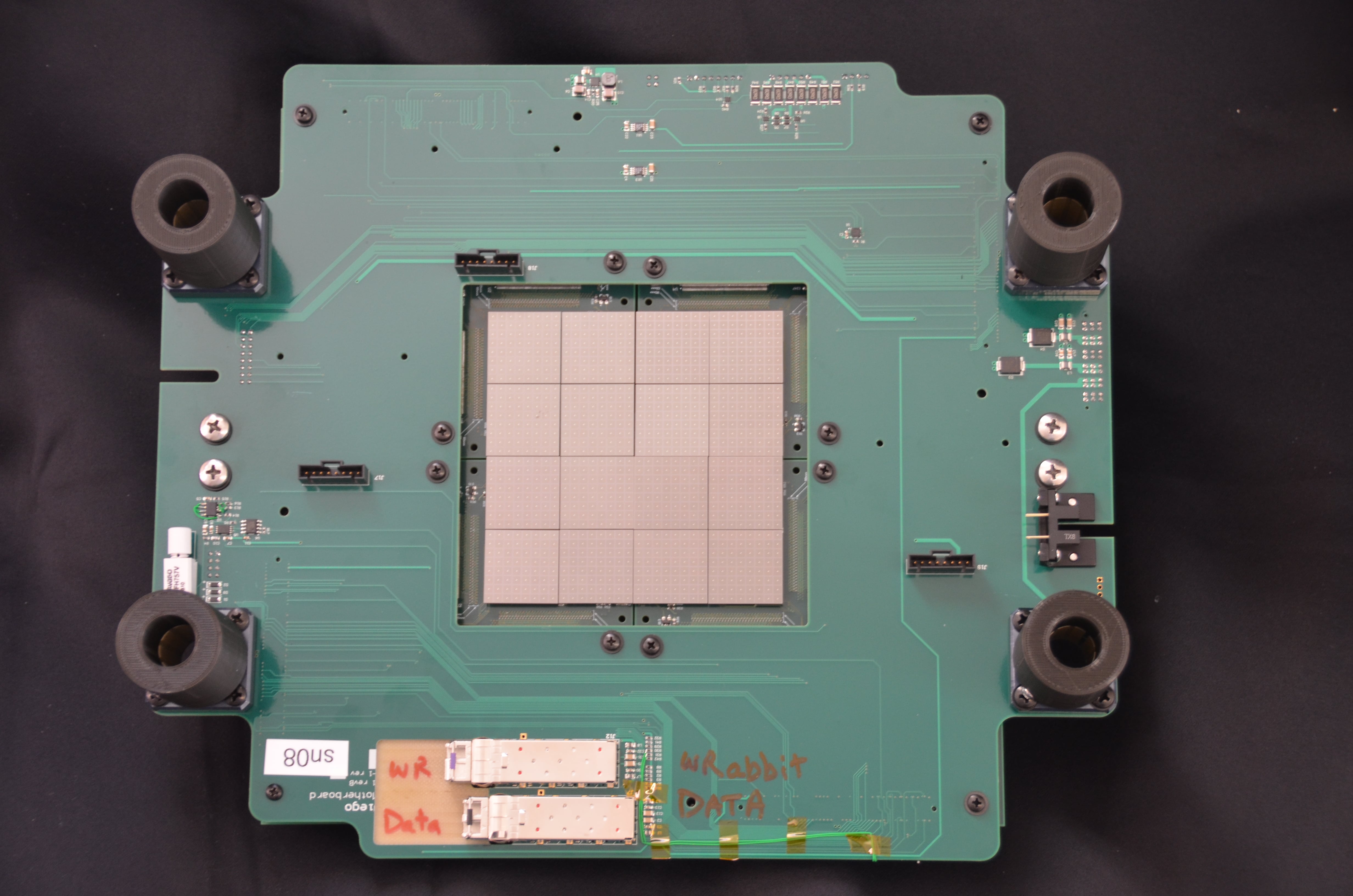}
		}
	\end{center}
	\caption{
		\label{photo_of_mobo}
		Photo of Mother Board, with four 256 pixel quadrant boards plugged into it, for a total of 1024 SiPM pixels.  
	}
\end{figure}

\indent Each motherboard (``mobo'') holds four quadrant boards, positioned with their 256-pixel SiPM corner-located arrays abutted, creating a 1024-pixel array that is 100\,mm square. One of the quadrant boards is the master, which talks to the host computer through fiber directly, including receiving commands from the host computer and sending scientific data and housekeeping data to the host computer. The other three quadrant boards are connected as a chain, with data received and sent through the master. The master quadrant board is also working as a White Rabbit slave, which is synced to a WR master through a fiber. The white rabbit derived 10\,MHz is sent out from the master quadrant board to a clock distributor, and then distributed to all the four quadrant boards. To get the precise timing, the trace lengths from the clock distributor to the quadrant boards are matched. 1\,PPS also derives from the WR slave, and distributed to the four quadrant boards. \\
\indent Some mechanical control interfaces are also designed on the Mother Board: there's a shutter control interface, to open or close the shutter, with shutter status to confirm the shutter position; there's also an LED pulse generator, to generate a flash for system verification. A block diagram of the mother board is shown in Figure~\ref{sch_of_mobo}, and a photo in \ref{photo_of_mobo}.\\

\subsection{Power System}
\indent The electronic system needs a set of DC voltages, shown in Table \ref{power_table}. We will have about 45 telescopes in each dome, so it's necessary to have a stable and compact power system. 
\begin{table}[ht]
	\caption{DC voltages for the system}
	\label{power_table}
	\begin{center}
		\begin{tabular}{|p{8cm}<{\centering}|p{5cm}<{\centering}|}
			\hline									
			\textbf{Usage} 							& \textbf{Voltage(V)}  \\
			\hline
			SiPM Bias Voltage 						& $-$70  \\
			\hline
			FPGA Core Power 						& 1.0  \\
			\hline
			FPGA GTX transceiver Power 				& 1.2  \\
			\hline
			FPGA Core Power 						& 1.8  \\
			\hline
			Power for Maroc3, external ADCs etc. 	& 3.3  \\
			\hline
		\end{tabular}
	\end{center}
\end{table}

\begin{figure}[htbp]
	\begin{center}
		\includegraphics[height=6.5cm]{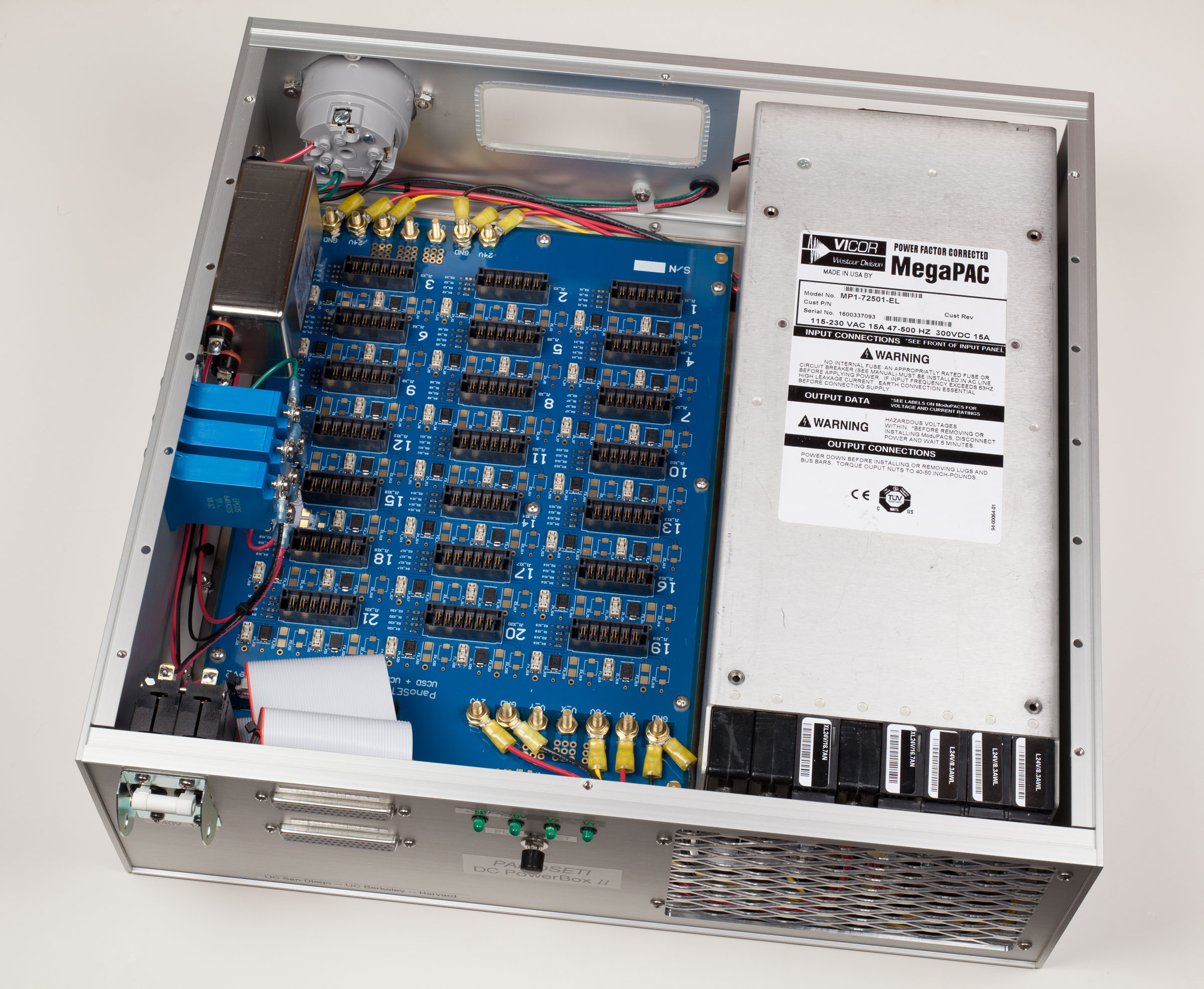}
		\caption{
			\label{photo_of_power}
			Rack-mount DC power supply for 21 telescopes.
		}
	\end{center}
\end{figure}

\indent Considering all the power requirements, we designed a on-board power system and a outboard power system. The on-board power system on mother board and quadrant board consists of some DC-DC and low dropout regulators(LDO), which generate different DC voltages for the electronic system, such as 1V and 1.8V for FPGA core power, 1.2V for the FPGA GTX transceiver, and 3.3V for the Maroc3 chips and external ADC chips. These LDOs can convert an input voltage to the needed voltages. The outboard power system is housed in 3U racks with low-noise switchmode DC converter modules (Fig.~\ref{photo_of_power}), able to power up to 21 modules with $-$70\,V for the SiPMs, and +24\,V for DC-DC and LDOs on the mother boards and quadrant boards. The DC voltage power cable has four wires (+24V/$-$70V/GND/GND) to each telescope.

\section{PANOSETI Gateware}
\begin{figure}[htbp]
	\begin{center}
		\includegraphics[height=5cm]{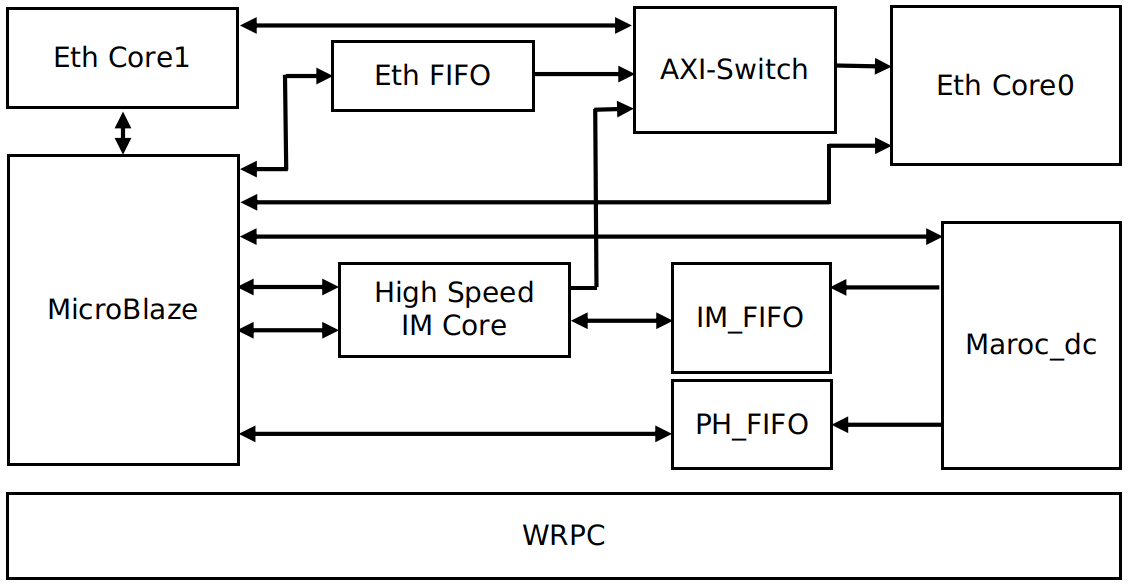}
		\caption{
			\label{block_diagram_of_gateware}
			PANOSETI FPGA gateware design:  The Maroc\_dc core is used for getting digital data from Maroc ASIC chip; then the data is transferred to IM\_FIFO and PH\_FIFO for different observation modes\ref{PH_mode},\ref{IM_mode}. Continuous imaging mode will generate images at a high frame rate (maximum 100,000 frames/s), so the packets are generated in the High Speed IM core in hardware. Pulse height packets will be generated in the Microblaze core at lower frame rate. Two Ethernet cores are in the gateware for the data chain from the four quadrant boards to computers.
		}
	\end{center}
\end{figure}

\indent In the FPGA gateware, we have currently implemented two observation modes: Continuous Imaging (CIM) mode and Pulse Height (PH) mode. Pulse height mode is optimized for detecting optical pulse widths $<$30\,ns\cite{2018Panoramic}. Continuous imaging mode employs counters on every pixel that count over-threshold events to produce
images at a programmable frame rate. This mode is more sensitive to transients of 10$\mu$s or longer. White rabbit is also implemented in the gateware for precise timing. The block diagram of the gateware is shown in Figure \ref{block_diagram_of_gateware}.

\subsection{Pulse Height Mode}
\label{PH_mode}
\indent Pulse height mode, which has been standard for previous optical SETI experiments\cite{howard2004search,horowitz2001targeted,werthimer2001berkeley,wright2001improved, wright2014near}, is most sensistive to transients with pulse width less than 30\,ns. When any one of the 256 triggers from a Maroc3 chip exceeds the programmable threshold, pulse height data is sent to the host computer over a 1\,GbE port. Because over-threshold events will not occur frequently, pulse-height data only requires a modest data rate. so Ethernet packets can be formed by software rather that gateware.  Pulse height data is transmitted from the Maroc\_dc core by FPGA gateware, and then packetized and sent out from the microblaze Xilinx soft processor, running at 100\,MHz. \\

\subsection{Continuous Imaging Mode}
\label{IM_mode}
\begin{figure}[htbp]
	\begin{center}
		\includegraphics[height=6cm]{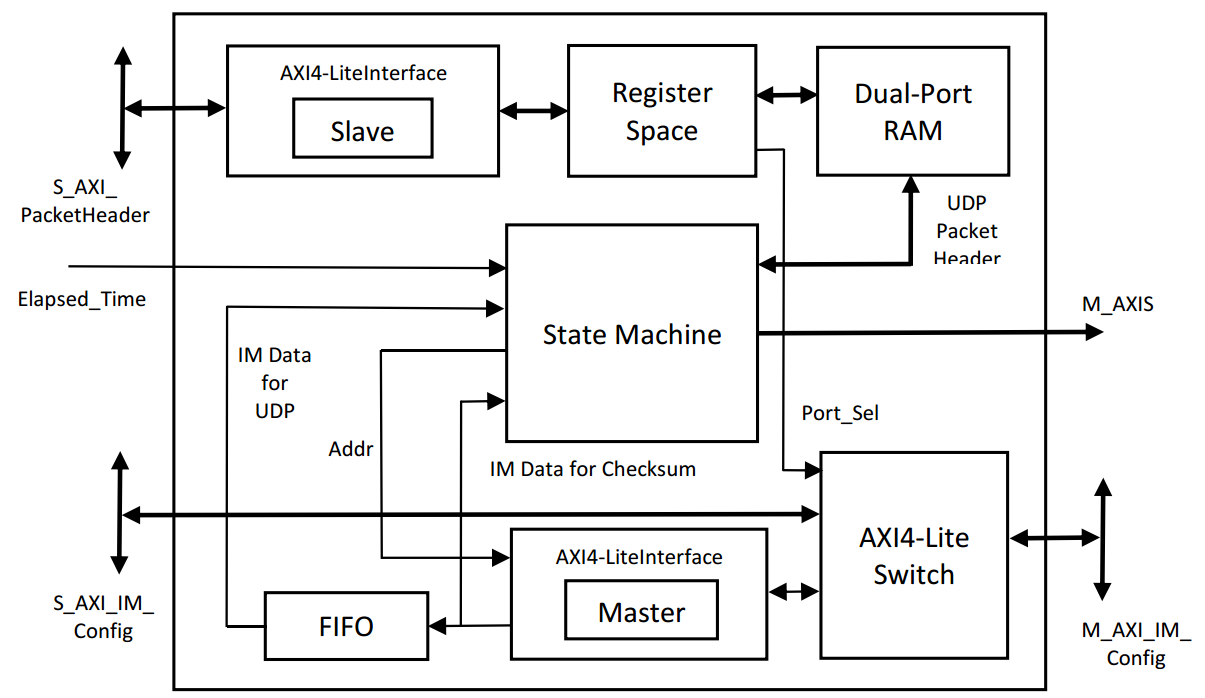}
		\caption{
			\label{block_diagram_of_hs_im}
			FPGA high speed imaging block diagram:  Here are three AXI4 interfaces on the HS-IM Core. S\_AXI\_PacketHeader is used for writing packet header information to the Dual-Port RAM. S\_AXI\_IM config is used for configuring IM\_FIFO from Microblaze core. M\_AXI\_IM Config is used for getting data from IM\_FIFO. The State machine in HS-IM core is the key part, which is used for generating image packets at a high frame rate. 
		}
	\end{center}
\end{figure}

\indent Continuous imaging mode employs counters on every pixel that count over-threshold events to produce
images at a programmable frame rate, which could range from 100\,Hz to 100\,kHz. But the microblaze core is too slow for the highest rates, so all imaging packets are generated in a high speed image mode (HS-IM) core in FPGA gateware, which gets data from the imaging FIFO, and generates imaging packets in hardware. The block diagram of the HS-IM core is shown in Figure \ref{block_diagram_of_hs_im}.

\indent Before continuous imaging mode is started, packet header data is written into the dual-port RAM, such as source/destination MAC addresses, source/destination IP address, and so on. When continuous imaging mode is running, the state machine gets packet header data and writes it to a packet FIFO through the M\_AXI interface first, and then writes the current elapsed time to the packet FIFO. The state machine gets image data from the image FIFO through the M\_AXI\_IM\_Config interface, and writes the data to the packet FIFO. The image packet is generated and sent to the host computer over the 1\,GbE port.

\subsection{White Rabbit}
	\label{sec:white_rabbit}
\indent For pulse height measurements, we need to know the precise time of each received optical pulse, with differential timing accuracy between the two domes ideally no more than a nanosecond. Conventional time distribution uses a GPS receiver, synchronized via satellite, with its 10\,MHz and 1\,PPS outputs distributed by transmission lines to all telescopes in a dome. \\
\indent This method is simple, requiring only signal buffers and dozens of coaxial lines. But the cable lengths must be matched to minimize timing skew.  Moreover, it requires a pair of coaxial lines per telescope module.  An elegant solution is the White Rabbit protocol.\\
\indent White Rabbit creates an Ethernet-based network with low-latency, deterministic packet delivery, and network-wide, transparent, high-accuracy timing distribution\cite{2011WhiteRabbit}. With Layer-1 synchronization, Precise Time Protocol (PTP) and precise phase measurements technologies, White Rabbit can achieve 1\,PPS time accuracy better than 1\,ns, and time precision better than 50\,ps\cite{daniluk2012white} through a fiber. White rabbit can measure and compensate for the delay from different length of fibers automatically, so the time accuracy and precision is stable, and independent of fiber lengths. There are three nodes are in WR network: Grandmaster, Master and Slave.

\begin{itemize}
	\item GrandMaster : synchronized to an external 1\,PPS and 10\,MHz reference source.
	\item Master      : transfer precise timing to other WR-compliant devices.
	\item Slave       : synchronizes its internal oscillator to another WR Master device. 
\end{itemize}

\indent In the white rabbit network, we have a White Rabbit switch working as grandmaster, which is provided with 1\,PPS and 10\,MHz from a GPS disciplined time/frequency reference. The hardware for White Rabbit (\S\ref{sec:quabo}) implements the White Rabbit PTP Core (WRPC) in gateware, which means each quadrant board is working as a pulse signal processor and White Rabbit slave simultaneously. Therefore, all the telescopes are White Rabbit slaves, connected to the White Rabbit switches through fibers ranging from a few meters to a kilometer. The resulting timing system is both compact and nanosecond-precise, and, being fiber based, does not require copper connections.

\section{Networking}
	\label{sec:networking}
\indent In PANOSETI project, $\sim$90 telescopes will be deployed between two domes for covering $\sim$4,441 square degrees. A large quantity of scientific packets will be generated over the nightly observations, sent via a data network to storage and computing units. We also need a computer cluster for controlling and monitoring the domes and all the telescopes, and processing and storing the scientific data.

\subsection{PANOSETI Network}
\begin{figure}[htbp]
	\begin{center}
		\includegraphics[height=8cm]{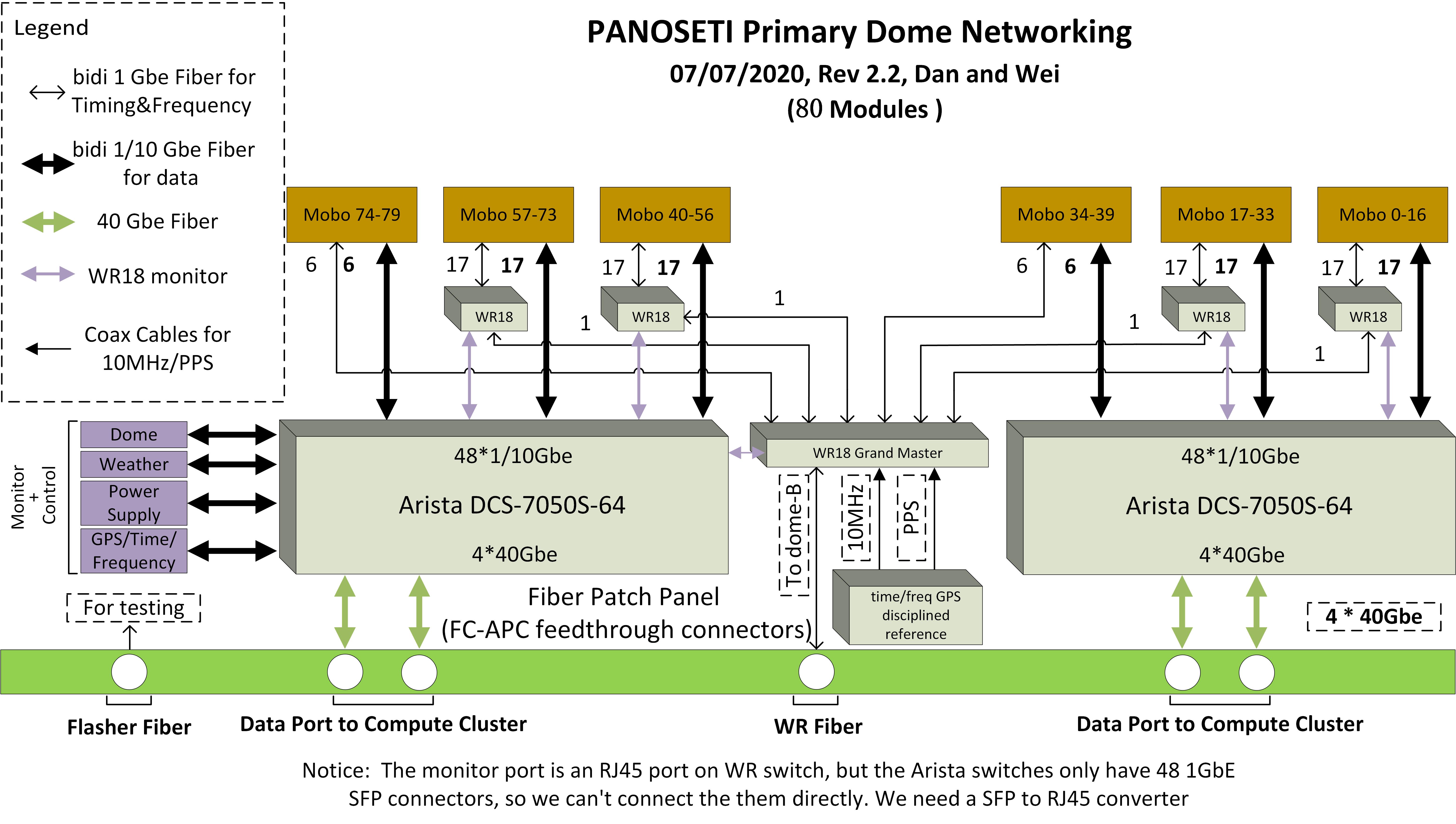}
		\caption{
			\label{primary_dome_network}
			PANOSETI Primary Dome Networking
		}
	\end{center}
\end{figure}

\indent Two domes will be deployed at each PANOSETI observatory. One of them is called the primary dome, in which we install telescopes, computer cluster, GPS receiver, some White Rabbit switches, and some data switches. The other is the secondary dome, in which we only install telescopes and data switches. The network design for the primary dome is shown in Figure~\ref{primary_dome_network}.

\indent The dome has two Arista DCS-7050S-64 network switches, each one has 48 1/10\,GbE ports and 4 40\,GbE ports. The 40 1\,Gbps(max) data streams are connected to each switch, so the total data stream can be transferred out from four 40Gbps port to the computer cluster. \\
\indent We also have five White Rabbit switches (WRS), which creates an independent White Rabbit network for timing synchronization. One of them is working as grandmaster, which connects to the GPS receiver and other four WRS. The other four WRS are working as masters, which is synced to grandmaster, and transfer timing data to the 45 slaves (telescopes). Each WRS has a management Ethernet port for configuration and status monitoring. The 5 Ethernet ports on WRS are also connected to Arista switches.\\

\indent The network design for the secondary dome is almost the same as the primary one. The only difference is that there is no computer cluster or GPS receiver in this dome. The scientific data, housekeeping data, and timing data are transferred through kilometer long fibers between the two domes. 

\subsection{Computer Cluster}
\begin{figure}[htbp]
	\begin{center}
		\includegraphics[height=8cm]{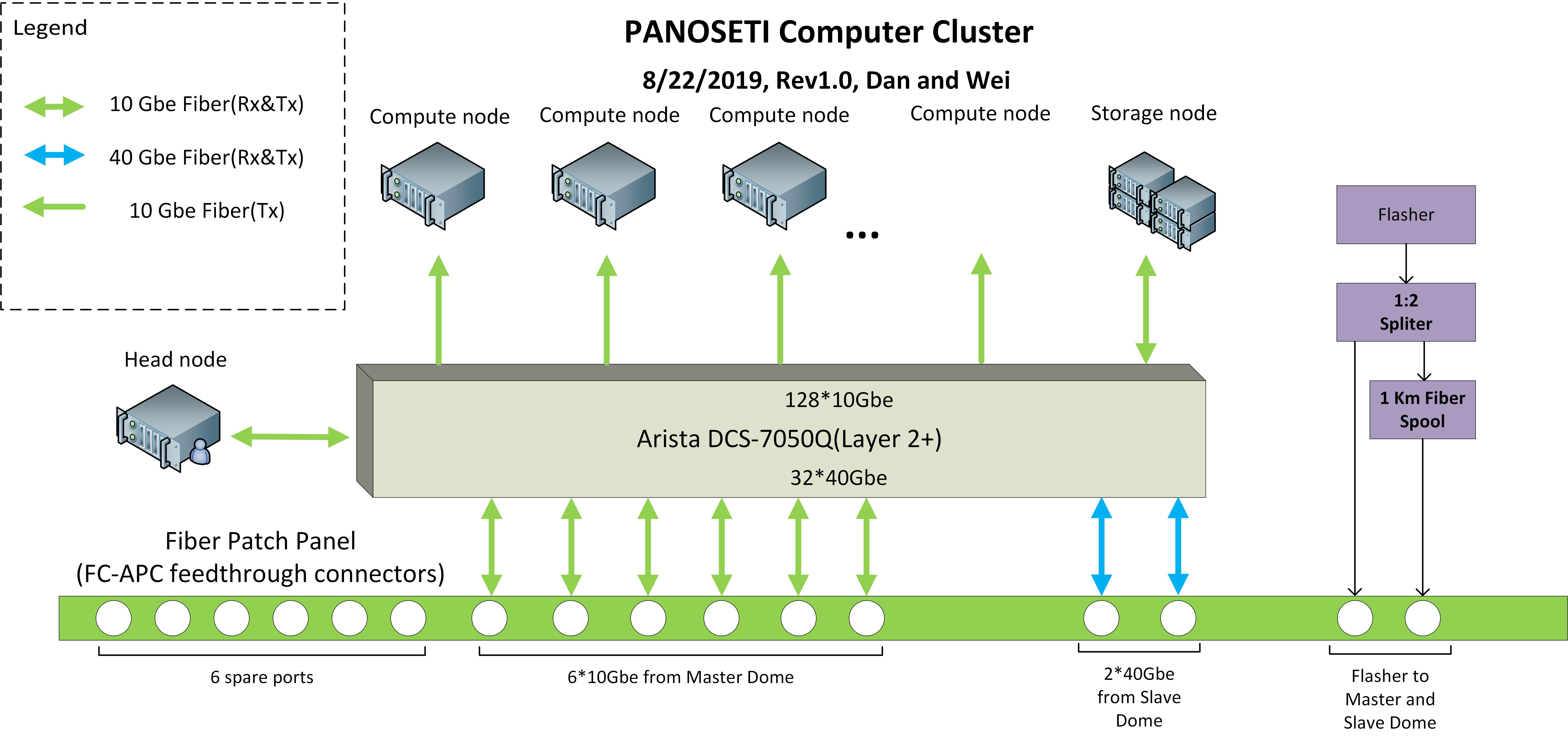}
		\caption{
			\label{computer_cluster}
			PANOSETI Computer Cluster
		}
	\end{center}
\end{figure}

\indent The computer cluster in the primary dome consists of a head-node computer, several compute nodes, and storage nodes. The head-node computer is the control unit, through which we can control both of the domes and telescopes, and monitor the status of the entire system, such as temperature, voltage, and so on. The compute node is used for data processing. The full data rate of 160\,Gbps is too much to store, so we use compute nodes to do some initial real-time data processing, such as spatial and time coincidence searching, so we only need to store potentially interesting events.  The storage nodes store the processed data in HDF5 files \ref{sec:hdf5} for subsequent off-line analysis. \\
\indent All of the computers are connected to a Arista DCS-7050Q(Layer 2+) switch, which has 128 10\,GbE ports and 32 40\,GbE ports. The 10\,GbE ports are connected to the head node, compute nodes, and storage nodes, and the 40\,GbE ports are connected to the primary and secondary domes for high speed data transmission. The computer cluster is shown in Figure \ref{computer_cluster}. \\

\section{Software}
	\label{sec:hdf5}
\indent PANOSETI archives data on imaging and pulse height events that are spatially- and time-coincident between two domes, all high SNR imaging or pulse height events, all imaging data at 781 Hz frame rate (1.28 ms integration time), as well as observatory and instrumental metadata. This data is stored in files written using a hierarchical data format, HDF5.
\subsection{What is HDF5}
\indent HDF5 (Hierarchical Data Format version 5) is an open source data file format that is specifically designed to store large complex heterogeneous data\cite{price2015hdfits}. Traditionally astronomical data have been stored using the FITS (Flexible Image Transport System) file format.  Although FITS has been widely used and supported, it lacks many features that HDF5 provides, such as heirarchical access, rapid I/O rates, and data compression.  More information regarding HDF5 can be found on The HDF Group's website, which is also where you are able to download the latest C library. HDF5 provides support for other languages, including python, which are wrappers for the C API.
\subsection{Software Tools Overview}
\indent PANOSETI uses many software tools such as HASHPIPE (High Availability Shared Pipeline Engine), Redis, InfluxDB. Its main data processing tool is HASHPIPE (\url{https://casper.ssl.berkeley.edu/wiki/HASHPIPE}), developed at Berkeley by Dave MacMachon, which allows for the acquisition of approximately 50 gigabits per second. HASHPIPE contains a structure of circular buffers to allow for fast acquisition of this large amount of data to become available for processing.
\begin{figure}
    \centering
    \includegraphics[width=\columnwidth]{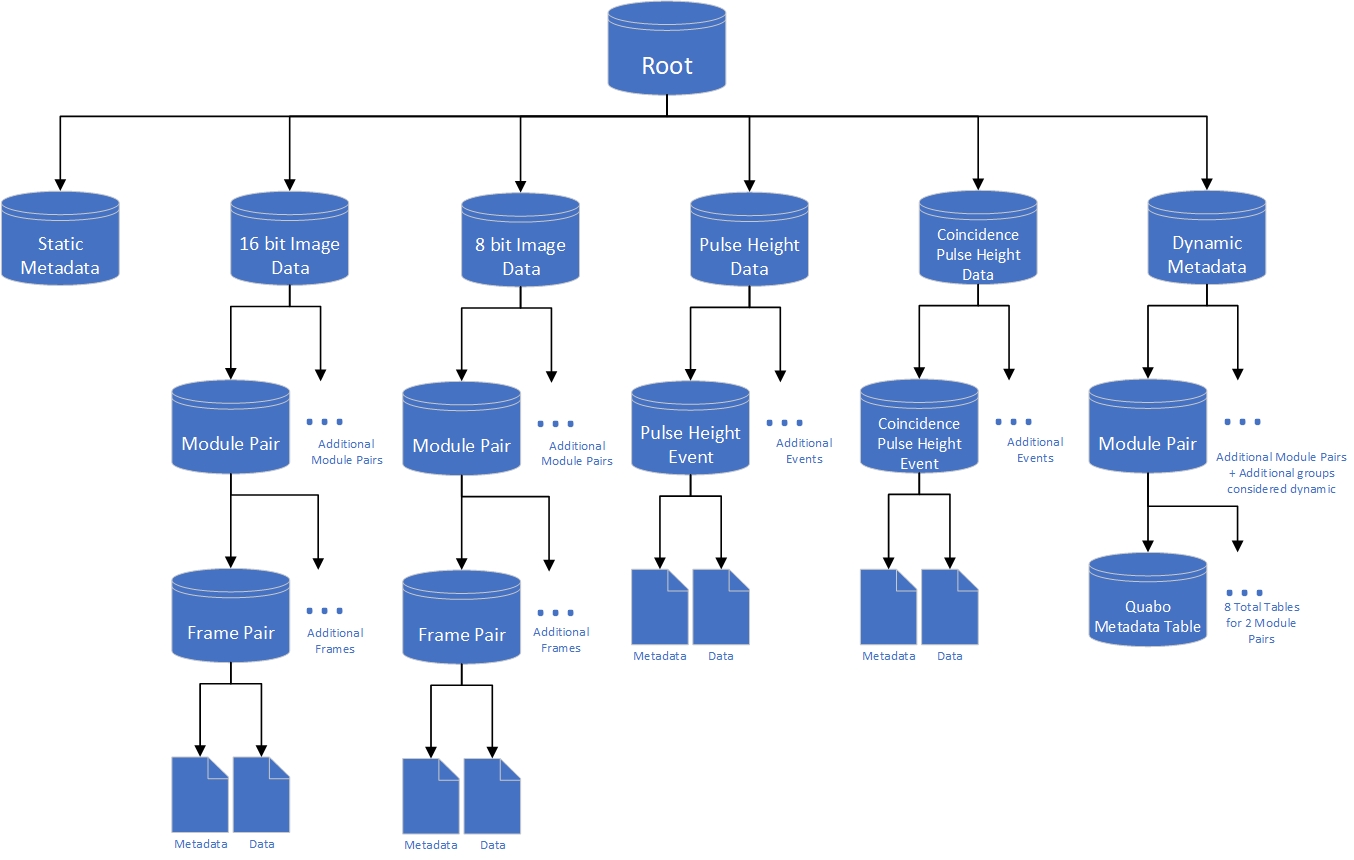}
    \caption{File Structure Diagram}
    \label{fig:file_structure}
\end{figure}

\subsection{Data File Structure}
\indent  Using HDF5's hierarchical data format we are able to categorize our data into groups. PANOSETI gathers two main data types which are imaging data, and pulse height data. Imaging data are images of the full sky at a 1.28 ms cadence given by integrating 64 20 us frames, and pulse height data are triggered events given some set threshold. Static metadata stores metadata, which are created once per file, and dynamic metadata are updated frequently through observations. PANOSETI updates metadata through a decentralized system where multiple scripts gather metadata from many sources at difference cadences, and pipes this data into a centralized REDIS and InFluxDB databases, as well as written to HDF5 file with the science data. The HDF5 data structure is shown in Figure~\ref{fig:file_structure}.

\section{Testing Results}
\begin{figure}[htbp]
	\begin{center}
		\includegraphics[height=8cm]{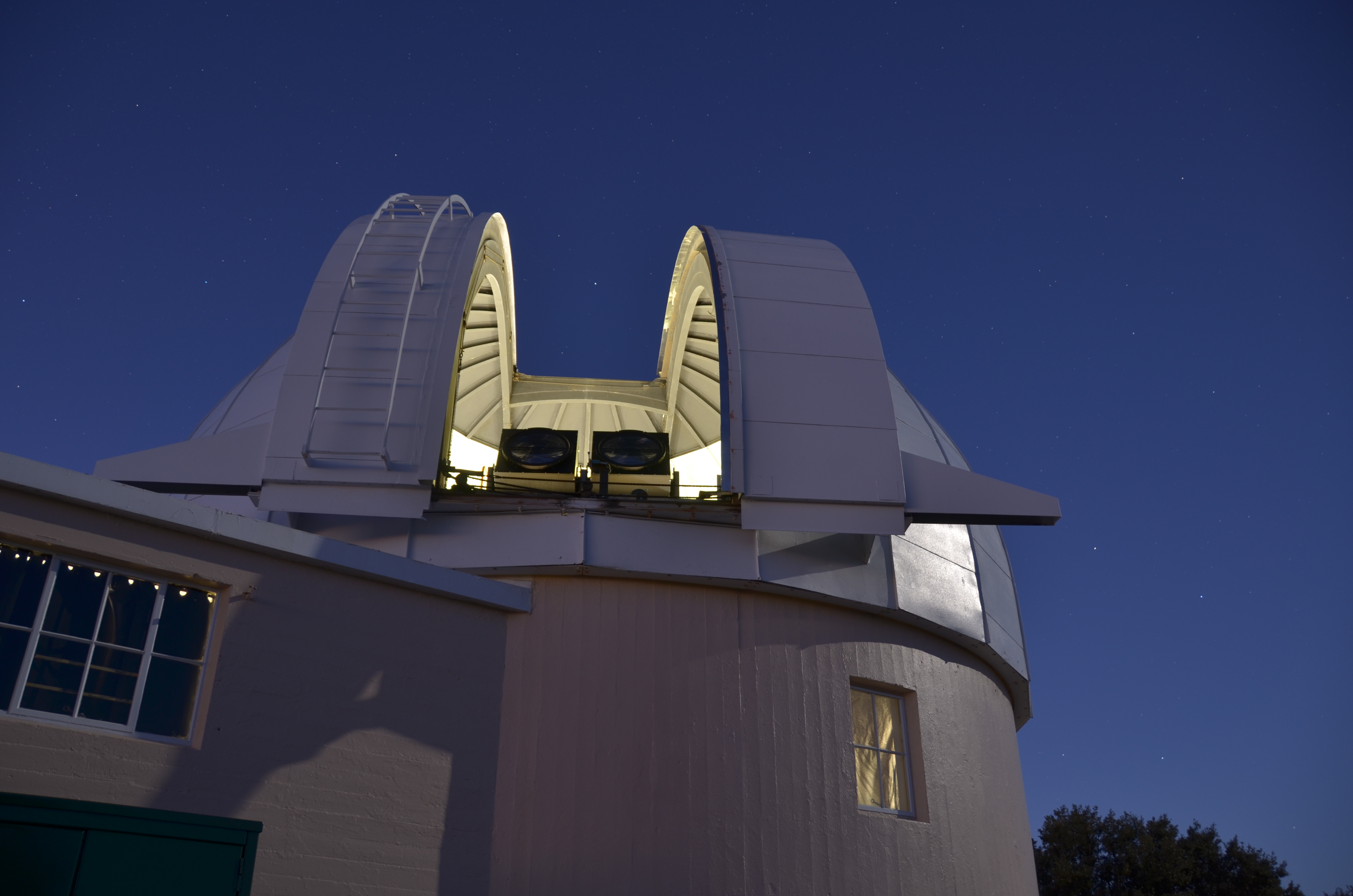}
		\caption{
			\label{telescopes_at_lick}
			PANOSETI Telescopes at Lick Observatory
		}
	\end{center}
\end{figure}

\indent Two telescopes, one data switch, one White Rabbit switch, and a computer have been deployed at Lick Observatory\ref{telescopes_at_lick} for prototype tests beginning in February 2020. The test results are provided in another paper in these proceedings (Maire et al. 2020 SPIE). \\

\section{Future Work}
\indent The PANOSETI project aims to search a large fraction of the visible sky for pulses at optical and near-infrared wavelengths, exploiting the efficiency and speed of solid-state photomultipliers. Later we plan to extend the response to near-infrared wavelengths\cite{2018Panoramic} in the same domes.

\section{Conclusions}
\indent The PANOSETI experiment will deploy observatory domes at several sites; each dome contains $\sim$45 telescopes and covers 4,441 square degrees. The focal plane electronics for each of the visible wavelength telescopes contains 1024 SiPM detectors, sensitive to 300--850\,nm wavelengths, feeding 1024 channels of analog and digital electronics in Maroc3 ASICs. With the electronic components on a mother board with four quadrant boards, we implemented continuous imaging mode and pulse height mode for capturing transients with pulse widths from nanoseconds to milliseconds, and with ns-accurate time stamping  via the White Rabbit protocol. The PANOSETI data network -- with switches, compute servers, and data storage servers -- is designed to accommodate the challenging rates of data transmission, real time data processing, and data storage. 

\acknowledgments      
\indent The PANOSETI research and instrumentation program is made possible by the enthusiastic support and interest of Franklin Antonio. We thank the Bloomfield Family Foundation for supporting SETI research at UC San Diego in the CASS Optical and Infrared Laboratory. Harvard SETI was supported by The Bosack/Kruger Charitable Foundation and The Planetary Society. UC Berkeley’s SETI efforts involved with PANOSETI are supported by NSF grant 1407804, the Breakthrough Prize Foundation, and the Marilyn and Watson Alberts SETI Chair fund. We would like to thank the staff at Mt.~Laguna and Lick Observatory for their help with equipment testing. The following vendors have been most gracious with their answers to our frequent technical questions: Hamamatsu, Weeroc, CAEN Technologies, and Amplification Technologies.

\bibliography{report} 
\bibliographystyle{spiebib} 

\end{document}